\documentclass[conference]{IEEEtran}

\def\BibTeX{{\rm B\kern-.05em{\sc i\kern-.025em b}\kern-.08em
    T\kern-.1667em\lower.7ex\hbox{E}\kern-.125emX}}

\usepackage{cite}
\usepackage{amsmath,amssymb,amsfonts}
\usepackage{algorithmic}
\usepackage{graphicx}
\usepackage{textcomp}
\usepackage{xcolor}
\usepackage{makecell}

\usepackage{siunitx}
\usepackage{hyperref}
\usepackage[nolist]{acronym}
\usepackage[flushleft]{threeparttable}
\usepackage{placeins}
\usepackage{svg}                                   
\usepackage[percent]{overpic}
\usepackage[nameinlink, capitalise]{cleveref}

\usepackage{mathtools}
\usepackage{booktabs}
\usepackage[nolist]{acronym}
\usepackage{pbalance} 
\usepackage[export]{adjustbox}

\usepackage{eso-pic}    

\usepackage{orcidlink}

\usepackage{enumitem}

\usepackage{comment}

\usepackage{scalerel}

\usepackage{booktabs}
\usepackage{tabularx}
\usepackage{makecell}
\usepackage{booktabs}
\usepackage{array}

\newcommand{\change}[1]{{\textcolor{black}{#1}}}       

\newcommand{\RNum}[1]{\uppercase\expandafter{\romannumeral #1\relax}}   

\DeclareRobustCommand{\IEEEauthorrefmark}[1]{\smash{\textsuperscript{\footnotesize #1}}}

\DeclareSIUnit\db{dB}                           
\DeclareSIUnit\dbv{dBV}                           
\DeclareSIUnit\dbi{dBi}                         
\DeclareSIUnit\dbm{dBm}                         
\DeclareSIUnit\dbu{dBu}                         
\DeclareSIUnit\watthour{Wh}                     
\DeclareSIUnit\mbps{Mbps}                       
\DeclareSIUnit\kbps{kbps}                       
\DeclareSIUnit\bps{bps}                         
\DeclareSIUnit\fps{fps}                         
\DeclareSIUnit\mAh{mAh}                         
\DeclareSIUnit\msInference{ms/inference}        
\DeclareSIUnit\persquare{\ensuremath{/\square}}

\begin{document}
\bstctlcite{IEEEexample:BSTcontrol} 

\AddToShipoutPictureBG*{
  \AtPageUpperLeft{%
    \put(0,-40){\raisebox{15pt}{\makebox[\paperwidth]{\begin{minipage}{21cm}\centering
      \textcolor{gray}{This article has been accepted for publication in the proceedings of the \\
       International Symposium on Sensor Applications (SAS 2026)\\ 
       } 
     \end{minipage}}}}%
   }
   \AtPageLowerLeft{%
     \raisebox{25pt}{\makebox[\paperwidth]{\begin{minipage}{21cm}\centering
       \textcolor{gray}{ \copyright 2026  Authors and IEEE. 
        This is the author’s version of the work. It is posted here for your personal use. Not for redistribution. \\
        The definitive Version of Record will be published in the proceedings of the International Symposium on Sensor Applications (SAS 2026).
         }
     \end{minipage}}}%
   }
 }


\title{A Time-Synchronized Video Reference System for Data Analysis of Body-Attached Sensor Nodes in Outdoor Scenarios
}

\author{
    \IEEEauthorblockN{
        Lukas Schulthess\;\orcidlink{0000-0002-6027-2927}\,\IEEEauthorrefmark{1},
        Fabian Pleisch\,\IEEEauthorrefmark{1},
        Matheo Käch\,\IEEEauthorrefmark{3}\\
        Björn P. Bruhin\,\IEEEauthorrefmark{2}\textsuperscript{,}\IEEEauthorrefmark{3},
        Michele Magno\;\orcidlink{0000-0003-0368-8923}\,\IEEEauthorrefmark{1},
        Luca Benini\;\orcidlink{0000-0001-8068-3806}\,\IEEEauthorrefmark{1},
        Christoph Leitner\;\orcidlink{0000-0002-7058-7236}\,\IEEEauthorrefmark{1}
        }
    \vspace{1mm}
    \IEEEauthorblockA{
        \IEEEauthorrefmark{1}\,\textit{Department of Information Technology and Electrical Engineering, ETH Zurich, Zurich, Switzerland}\\
        \IEEEauthorrefmark{2}\,\textit{Swiss Federal Institute of Sport Magglingen, Magglingen, Switzerland} \\
        \IEEEauthorrefmark{3}\,\textit{Swiss-Ski, Worblaufen, Switzerland}
        }
}

\maketitle

\begin{abstract}

Wearable body-attached multi-sensor systems enable detailed analysis of human motion and physiological signals in sports, rehabilitation, and movement research. 
While wireless synchronization techniques can reliably align sensor data streams, interpreting and validating complex or unconstrained activities often requires an additional, objective visual reference.
Existing laboratory-grade reference systems provide high accuracy but are impractical for outdoor or field deployments.
\change{In contrast, commercial video timecode solutions typically rely on local device-to-device synchronization, which increases the power required to maintain synchronization. This is not desirable in many application scenarios.}
This paper presents a lightweight \ac{tcg} that converts \ac{gnss}-derived time directly into a \ac{ltc} signal that is injected into the recording via a camera audio channel. The approach eliminates continuous handshaking, allowing the system to be activated immediately before the action of interest, thus reducing power consumption and enabling smaller batteries and unobtrusive hardware designs of body-attached sensor nodes. The \ac{tcg} supports common video frame rates of 24, 25, and 30\,\ac{fps}. Experimental evaluation confirms that accurate time alignment is maintained for several minutes without \ac{gnss} updates. At 30\,fps, the alignment duration is 543\,s before a potential frame-level shift occurs. With an average power consumption of 35.37\,mW, the system achieves an operating time of up to 75\,h when powered by two standard AAA alkaline batteries.

\end{abstract}

\vspace{5pt}

\section{Introduction}\label{sec:introduction}

\change{Wearable, body-attached} multi-sensor systems have become a key technology for capturing human motion and physiological signals. 
By combining data from multiple wearable sensing devices positioned at different body locations, such as wrist~\cite{j:li_multisensor_hand_gesture_2021}, knee~\cite{j:gu_multisensor_system_2023}, or torso~\cite{j:zaltieri_multisensor_sitting_position_2023}, in-depth situational or motion analysis is enabled.
This spatial distribution of sensing locations increases information density and improves situational awareness compared to single-device or single-location approaches. Such systems are widely used in sports performance analysis~\cite{j:swain_multisensor_sport_2023}, rehabilitation~\cite{j:salis_multisensor_rehabilitation_2023}, and general movement analysis~\cite{j:lu_limb_locomotion_phases_2021} for post-action data investigation and interpretation.
\begin{figure}[t]
    \centering
    \includegraphics[width=\columnwidth]{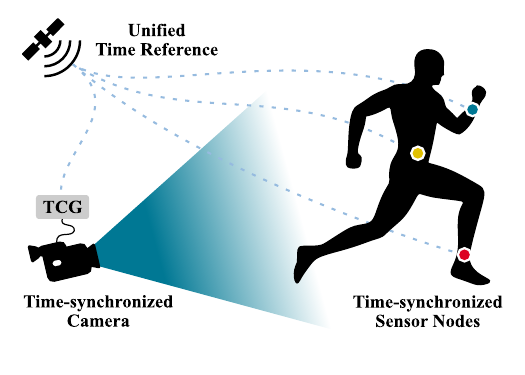}
    \vspace{-13mm}
    \caption{Multiple body-worn devices, time-synchronized via \ac{gnss}, record data during outdoor activities. To provide a visual ground truth and a temporal reference for post-event analysis using consumer-grade cameras, a dedicated Timecode Generator (TCG) has been developed to embed precise timing information into each recorded video frame.}
    \vspace{-6mm}
    \label{fig:ltc_timestamper_concept}
\end{figure}

However, while data collected from multiple body locations can already provide valuable insights, the full potential of multi-sensor systems is realized only when all sensor data streams are precisely aligned on a common time axis.
Accurate temporal synchronization enables joint analysis of multi-location sensor data to provide a holistic view of complex motion patterns.
To achieve this, \change{wearable sensor nodes are commonly synchronized between each other} using wireless short-range protocols such as \ac{ble}~\cite{j:kim_wireless_2025, j:ohara_wireless_timesync_2024, j:polonelli_wireless_time_synchronization_2023} or Enhanced ShockBurst~\cite{c:esb_krull_2025}, achieving synchronization accuracy below \qty{1}{\milli\second} by regularly exchanging synchronization packets. This ensures that dynamic events are accurately aligned across all sensor data streams and enable precise temporal correlation and on-device processing or post-action analyses~\cite{sensor_fusion_kulvicius_2025}. 

Despite synchronized sensing, data analysis remains challenging when relevant features are not clearly identifiable from the sensor data alone~\cite{movement_data_reference_hafner_2023}.
This is often the case in unconstrained environments, when activities are unknown or when motion sequences are complex. Interpreting sensor data in a movement context requires a camera-based visual ground-truth reference, which in turn necessitates accurate temporal \change{synchronization between the camera stream and sensor nodes}.
This can be achieved by embedding time information directly into the video stream. Using \ac{ltc}~\cite{SMPTEST12-1}, timing data can be injected via the camera’s audio channel.
Several commercial systems adopt this approach, with synchronization established through dedicated handshake procedures between devices ~\cite{tentacle_sync_e, atomos_ultrasync_blue, deity_tc1_timecode_box}.
In these setups, the unified time reference is typically obtained from a master device, often a smartphone, and distributed to all devices. Synchronization must then be maintained after the initial handshake throughout the recording period, which increases the system’s power consumption.

\change{In contrast, body-worn sensor systems must be compact, lightweight, and unobtrusive to avoid interfering with the wearer's natural movement, imposing strict limits on battery capacity.
In outdoor scenarios such as ski jumping, there can be a substantial delay between the initial synchronization handshake of the camera and sensor node and the onset of the activity of interest~\cite{c:skilog_schulthess_2023}.
Maintaining synchronization during this idle period drains battery resources, thereby reducing the system's operational lifetime~\cite{clock_drift_salimnejad_2025}.
Another way to achieve synchronization is to use a \ac{gnss}-based time reference instead of a device-to-device handshake. This approach eliminates the need for direct coordination between devices, as cameras and body-attached sensor nodes independently align their data streams using a common satellite-provided time source, as shown in \cref{fig:ltc_timestamper_concept}.}
\change{As a result, the camera and sensor nodes can be powered on immediately before the action of interest for synchronization with the \ac{gnss} time base, thus reducing the active time of all system components and overall power consumption.
This allows the use of smaller batteries and lighter hardware while improving usability in field deployments.}


\change{In this context, we present a portable, lightweight \ac{tcg} that generates \ac{ltc}-encoded timestamps directly from \ac{gnss}-derived satellite time, without requiring any external synchronization equipment. Combined with \ac{gnss}-synchronized body-attached sensor nodes, this system enables synchronized video and sensor data collection by relying on a common \ac{gnss}-based time source.}
In particular, this article presents the following contributions:

\begin{enumerate}
    \item \textbf{Timecode Generator:} A compact, standalone system that generates \ac{ltc} signals derived from \ac{gnss}-based time information, specifically tailored for the microphone inputs of consumer-grade video cameras.
    \item \textbf{Multi–frame-rate support:} User-selectable frame rates of 24, 25, and 30~\ac{fps}, with adjustable output amplitude in the range of \qtyrange{1}{50}{\milli\volt} to accommodate the microphone input requirements of consumer-grade cameras.
    \item \textbf{Timing Analysis:} A detailed timing analysis verifies the precise and correct generation of \ac{ltc} signals and quantifies the duration for which synchronization remains stable in the absence of \ac{gnss} updates. 
\end{enumerate}

\section{Background \& Related Work}\label{sec:related_work}

\begin{figure*}[t]
    \centering
    \begin{overpic}[width=\textwidth]{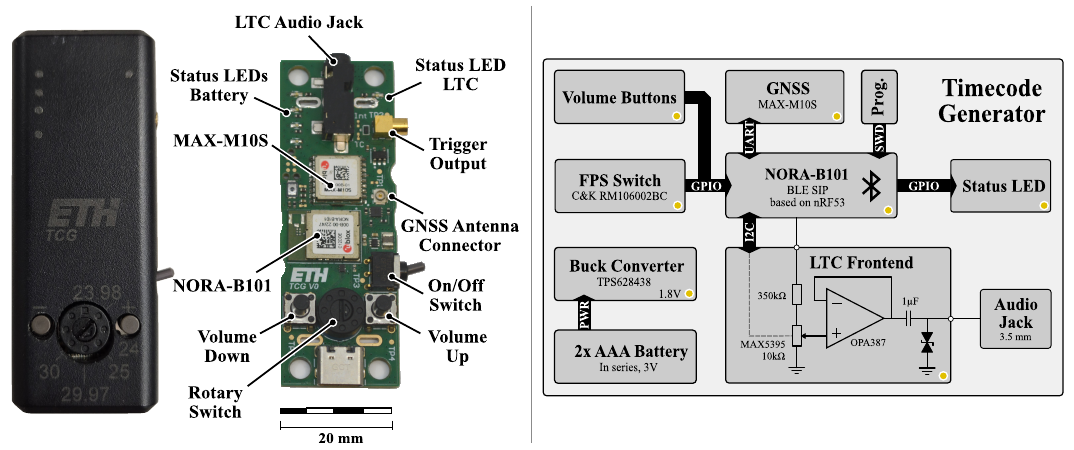}
        \put(1,40.5){(a)}
        \put(50.5,40.5){(b)}
    \end{overpic}
    \vspace{-8mm}
    \caption{System overview: (a) Top-view of enclosure and \ac{pcb} of the implemented timecode generator, (b) High-level block diagram including the detailed \ac{ltc} frontend sub-circuit.}
    \vspace{-6mm}
    \label{fig:tcg_system}
\end{figure*}


\acf{ltc} is a standardized and widely adopted method for time-stamping video recordings in professional audio-visual production and broadcast systems~\cite{SMPTEST12-1}. It encodes absolute time information as an audio signal that is recorded alongside the video on a conventional audio channel, typically injected via the microphone input. By embedding timing information directly into the audio stream, \ac{ltc} enables time recovery of hour, minute, second, and frame number information for each video frame during post-processing.

During post-processing, the encoded timestamps can be extracted from the audio stream using dedicated software, such as the open-source library \textit{libLTC}~\cite{libltc}. Owing to its simplicity, robustness, and compatibility with consumer and mid-range commercial recording equipment, \ac{ltc} is well suited for time-synchronized multimedia acquisition and distributed measurement systems.
Each \ac{ltc} frame consists of a fixed-length 80-bit word.
Of these, 26 bits encode the timestamp in the format \textit{frame:second:minute:hour}. An additional 32 bits are reserved for user-defined data fields, referred to as binary groups, which may contain auxiliary metadata such as recording date or geographic information. The remaining bits are allocated to synchronization and control functionalities to ensure reliable decoding.

The \ac{ltc} signal is modulated using \ac{bmc}~\cite{b:springer_wireless_line_coding_2022}.
\ac{bmc} is a self-clocking line-coding scheme that ensures regular signal transitions for reliable clock recovery. A binary zero is represented by a single transition at the beginning of the bit period. A binary one is represented by two transitions, one at the beginning and one at the midpoint of the bit interval. Because \ac{ltc} is transmitted continuously, the resulting bit rate directly depends on the selected video frame rate.
For example, at 30~\ac{fps}, an 80-bit \ac{ltc} frame results in a bit rate of 2400\,bit/s. This corresponds to a signal frequency of \qty{2400}{\hertz} for binary zeros and \qty{4800}{\hertz} for binary ones.

A fundamental limitation of \ac{ltc} is its limited temporal alignment accuracy relative to the actual camera frame capture time. This limitation arises because the \ac{tcg} and the camera operate independently, without direct communication or hardware synchronization.
The \ac{tcg} outputs the \ac{ltc} signal continuously, independent of the camera’s internal frame capture cycle. During post-processing, decoding software typically assigns the nearest available time code to each video frame.
Consequently, the assigned timestamp may differ by up to half a frame period and still be considered synchronized~\cite{SMPTEST12-1}.

\section{Timecode Generator}\label{sec:methods}
The \ac{tcg} presented in this article comprises a custom hardware platform and dedicated firmware that implement \ac{gnss}-based time synchronization and \ac{ltc} signal generation via deterministic \ac{gpio} toggling.

\subsection{Hardware}
The \acl{tcg} is designed for outdoor field operation. It is powered by replaceable batteries to avoid long charging times during deployment.
The device has a total mass of \qty{53.9}{\gram}, including batteries, and dimensions of \(28\times66\times24\) mm. 
This compact form-factor allows integration with standard camera recording equipment. \cref{fig:tcg_system} shows the hardware realization and the corresponding high-level block diagram.

\subsubsection{Processing and GNSS Timing Core}
The system is based on a \textit{Nordic Semiconductor nRF5340} \ac{soc} featuring two ARM Cortex-M33 cores and \ac{ble} capability.
The \ac{mcu} is integrated into a \textit{u-blox NORA-B101} module and is responsible for creating the \ac{ltc} signal.
Absolute time in \ac{utc} is received via \ac{uart} from a \textit{u-blox MAX-M10S} \ac{gnss} module. In addition, the \ac{gnss} module provides a dedicated time-pulse output that enables precise alignment to within several nanoseconds of the start of each second \cite{ublox2011_timing}. A \textit{MOLEX 206560} flexible \ac{gnss} antenna is used for signal reception.

\subsubsection{LTC Output Stage}
The \ac{smpte} standard for \ac{ltc} signals~\cite{SMPTEST12-1} recommends \ac{ltc} signal amplitudes between \qtyrange{1}{2}{\volt} and specifies an allowable range of \qtyrange{0.5}{4.5}{\volt} for professional line-level audio. However, consumer-grade microphone inputs typically operate at significantly lower signal levels, with a nominal line-level of \qty{-38}{\dbv}~\cite{b:adorno_microphones_2022}. Therefore, a dedicated output stage is required to attenuate the \ac{mcu}-generated digital signal of \qty{1.8}{\volt} to a range between \qty{1}{\milli\volt} and \qty{50}{\milli\volt}. Signal attenuation is implemented using a voltage divider consisting of a \qty{350}{\kilo\ohm} resistor and an \textit{Analog Devices MAX5395} programmable digital potentiometer with an end-to-end resistance of \qty{10}{\kilo\ohm}, controlled via \ac{i2c}. The scaled signal is taken from the potentiometer's wiper terminal and buffered by a \textit{Texas Instruments OPA387} \ac{opamp} configured as a voltage follower. The signal is then AC-coupled through a \qty{1}{\micro\farad} X7R/50V ceramic capacitor. Assuming a minimum camera input impedance of \qty{1}{\kilo\ohm}~\cite{IEC60268-4}, this results in a \qty{-3}{\db} corner frequency of \qty{159}{\hertz} for the high-pass filter. 

The \ac{ltc} signal is output through a standard \qty{3.5}{\milli\meter} \ac{trs} audio jack, which is commonly used for microphone inputs.
On stereo-capable devices, the signal is typically present only on the left channel. The output amplitude is adjustable via two \ac{spst} control buttons.
A \textit{C\&K RM106002BC} six-position rotary switch allows selection of the target frame rate. An additional signal output, accessible via an \ac{mmcx} connector, provides a trigger signal at the start of every new second to synchronize external systems.

\subsubsection{Power Supply and Regulation}
The system is powered by two AAA alkaline batteries connected in series. The input voltage ranges from approximately \qty{3}{\volt} when fresh to \qty{1.9}{\volt} at the end of discharge. A \textit{Texas Instruments TPS628438} buck converter regulates the supply to a stable \qty{1.8}{\volt} system voltage.
Five \acp{led} provide operational feedback. Four \ac{led}s indicate the battery state-of-charge, whereas the status \ac{led} blinks at \qty{1}{\hertz}, indicating active \ac{ltc} signal generation.

\subsubsection{Mechanical Integration}
The circuit is implemented on a 4-layer \ac{pcb} with a thickness of \qty{1.6}{\milli\meter} using a standard layer stackup.
The board outline was designed to fit the \textit{New Age S3A-281109-K} ABS enclosure, which includes a dedicated compartment for two AAA alkaline batteries.

\subsection{Firmware}
The firmware is implemented using the Zephyr \ac{rtos}. It is designed as a fully interrupt-driven system.
All time-critical operations are handled by hardware timers operating at \qty{16}{\mega\hertz} and external interrupts.
No continuous main processing loop is required during normal operation, ensuring deterministic timing behavior and minimal software-induced jitter.

\subsubsection{Time Synchronization and System Timing}
\begin{figure}
    \centering
    \includegraphics[width=\columnwidth]{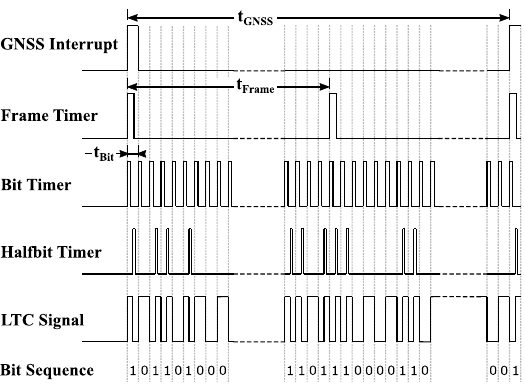}
    \vspace{-4mm}
    \caption{Timing diagram illustrating the generation of a \ac{bmc} \ac{ltc} signal. An external interrupt from the \ac{gnss} module initiates the sequence generation and provides precise time synchronization. The \textit{frame timer} defines the \ac{ltc} frame duration according to the user-selected frame rate. A \textit{bit timer} subdivides the frame into individual bit intervals, which are further subdivided by the \textit{half-bit timer} to support bi-phase mark encoding. Based on this timing hierarchy, the \ac{ltc} signal is generated by deterministic GPIO toggling. For each bit, a transition at the bit boundary is mandatory, while an additional transition at the half-bit boundary is inserted for logic ‘1’ symbols.}
    \vspace{-4mm}
    \label{fig:ltc_timer}
\end{figure}

Time synchronization is referenced to the \ac{gnss} module via its \textit{Timepulse} output, which marks the start of each \ac{utc} second with high temporal accuracy of \qty{60}{\nano\second}~\cite{ublox2025_maxm10s}. This interrupt serves as the global timing reference for the entire system. It triggers the generation of \ac{ltc} frames aligned to absolute time and resynchronizes the internal \(\pm\,30\,\)ppm \qty{32}{\mega\hertz} crystal of the \textit{u-blox NORA-B101} module when \ac{gnss} time synchronization is available. A hierarchical timer structure is used to generate the \ac{bmc} encoded \ac{ltc} signal by toggling a \ac{gpio}, as shown in \cref{fig:ltc_timer}.

\subsubsection{LTC Frame and Bit Timing Generation}
At the frame level, a dedicated \textit{frame timer} controls the start of each \ac{ltc} frame. The frame timer period is configured to match the video frame rate selected on the camera to ensure correct temporal alignment. The six-position rotary switch allows the user to select the desired frame-rate.
Before signal generation begins, the complete \ac{ltc} bit sequence for the upcoming frame is computed using the time information provided by the \ac{gnss} module.
Each frame is subdivided by a \textit{bit timer} that controls the timing of individual bits within the 80-bit \ac{ltc} word.
The bit timer toggles the \ac{ltc} output \ac{gpio} pin at the beginning of each bit interval to initiate the \ac{bmc}. If the transmitted bit is a logical ‘1’, the timer schedules an additional compare event at the midpoint of the bit period, referred to as \textit{half-bit} event, which produces the second transition required by the \ac{bmc} scheme. After all 80 bits of a frame have been transmitted, the bit timer is stopped and reset. It is restarted when the next frame timer event occurs.

\begin{table}
    \centering
    \renewcommand{\arraystretch}{1.2}
    \caption{Theoretical timing deviation introduced by the discretization of non-integer frame durations}
    \label{tab:timer_timing}
    \footnotesize
        \begin{tabularx}{\columnwidth}{@{}
        >{\centering\arraybackslash}p{0.2\columnwidth}
        >{\centering\arraybackslash}p{0.2\columnwidth}
        >{\centering\arraybackslash}p{0.2\columnwidth}
        >{\centering\arraybackslash}X
        @{}}
        \toprule
        \textbf{FPS} & \textbf{1/FPS} & \(\mathbf{t_{Frame}}\) & \textbf{Delay per Frame}  \\
        \midrule
        24 & $41.\overline{6}$\,ms   & 41.666625\,ms    & $-41.6$\,ns \\
        25 & 40.00\,ms              & 40.00\,ms        & 0 \\
        30 & $33.\overline{3}$\,ms  & 33.3333125\,ms   & $-20.8$\,ns \\
        \bottomrule
    \end{tabularx}
\end{table}

\subsection{System Characterization and Evaluation Procedure}
This section describes the experimental procedure used to assess the temporal accuracy of the generated \ac{ltc} signal.
Frame rates of 24, 25, and 30 \ac{fps} were analyzed to evaluate correct \ac{ltc} timestamp generation.
For this, a computer was synchronized to an ETH internal time \ac{ntp} server (time.ethz.ch) and used to display \ac{utc} time at a refresh rate of \qty{60}{\hertz}. Measurements were conducted using a \textit{Sony alpha 6400} camera with the exposure time set to 1/60 second and the \ac{tcg} connected to the camera’s microphone input via a standard \qty{3.5}{\milli\meter} audio cable. After the \ac{tcg} achieved time synchronization and began generating \ac{ltc} signals, indicated by a \qty{1}{\hertz} blinking of the status \ac{led}, video recording was started for one second, resulting in the same number of frames as the selected frame rate.
During post-processing, timestamps were extracted from the \ac{ltc} signal captured on the left audio channel and assigned to the corresponding video frames.
These extracted timestamps were then compared against the \ac{utc} time displayed within each frame. Temporal alignment was quantified using the mean mismatch, which represents the systematic timing offset, and the standard deviation, which characterizes temporal variability. In addition, the mean, the \ac{mae}, and the \ac{fsm}, defined as the temporal margin between the observed \ac{maxae} and the maximum allowable time shift, were computed to characterize the deviations.
Compared with the expected RMS timing uncertainty of the evaluation setup given in  \cref{eq:theoretical_deviation}, these results can be interpreted objectively.
%
\begin{equation}
  \sigma_{\text{total}} =
  \sqrt{
  \left(\frac{T_{\text{disp}}}{\sqrt{12}}\right)^2 +
  \left(\frac{T_{\text{exp}}}{\sqrt{12}}\right)^2
  }
  \label{eq:theoretical_deviation}
\end{equation}
%

For the equations above, it is assumed that the individual error contributions follow a uniform distribution.
$T_{\text{disp}}$ denotes the display refresh period and $T_{\text{exp}}$ the camera exposure time.

\subsection{Timing Analysis}\label{sec:timing_analysis}
To achieve the required temporal precision at bit level, the hardware timer is operated at its maximum frequency of \qty{16}{\mega\hertz}, corresponding to a timing resolution of \qty{62.5}{\nano\second} per timer tick. Depending on the selected video frame rate, this finite resolution introduces a deterministic timing deviation due to the discretization of non-integer frame durations. The resulting per-frame deviations are summarized in \cref{tab:timer_timing}.
In addition to discretization effects, timing accuracy is influenced by the frequency stability of the \qty{32}{\mega\hertz} crystal oscillator integrated in the \textit{u-blox NORA-B101} module. The crystal has a specified relative frequency accuracy of \(\pm\,30\,\)ppm. Furthermore, a fixed computational latency is introduced by the generation of the \ac{ltc} bit sequence for each frame. 
The maximum duration for which the system can remain synchronized without active \ac{gnss} time updates is defined as the point at which the accumulated timing error reaches half of a video frame period. This limit corresponds to the accepted temporal alignment bound of \ac{ltc}.
Considering all relevant timing contributions, the upper bound on uninterrupted synchronization time is computed using \cref{eq:tmax_upper_bound}:

\begin{equation}
\label{eq:tmax_upper_bound}
t_{\max} =
\frac{t_{\mathrm{half}} - t_{\mathrm{calc}}}
{\epsilon_{\mathrm{frame}} \cdot f_{\mathrm{fps}} + \delta_{\mathrm{ppm}}\cdot 10^{-6}}
\end{equation}

Here, $t_{\mathrm{calc}}$ denotes the fixed computation latency, $\epsilon_{\mathrm{frame}}$ represents the worst-case per-frame timing deviation due to timer discretization, $f_{\mathrm{fps}}$ is the video frame rate, and $\delta_{\mathrm{ppm}}$ denotes the relative frequency error of the crystal oscillator in parts per million.

\subsection{Power Analysis}\label{sec:timing_analysis}
Power measurements were conducted using the \textit{Nordic Semiconductor nRF Power Profiler Kit II}. 
To assess total system power consumption, the profiler operated in source-meter mode. 
For more granular analysis, the current of the active components has been measured using the profiler’s ampere meter mode
All measurements were performed at a constant voltage of \qty{1.8}{\volt}.
\section{Results}\label{sec:results}
\begin{figure}
    \centering
    \includegraphics[width=\columnwidth]{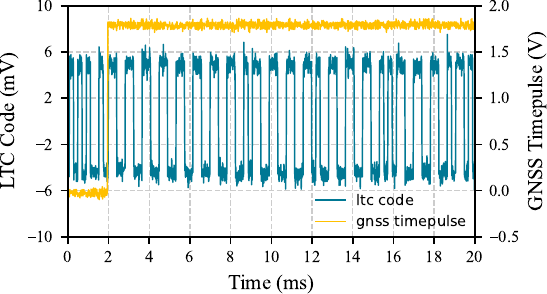}
    \vspace{-6mm}
    \caption{Sequence of the generated \ac{ltc} signal at 30 fps. The rising edge of each \ac{gnss} timepulse triggers a temporal realignment of the \ac{ltc} code,
    }
    \vspace{-6mm}
    \label{fig:ltc_signal}
\end{figure}

The resulting metrics for all evaluated frame-rates over N frames are summarized in \cref{tab:fps_test}.
A representative section of the continuous \ac{ltc} signal generated by the timecode generator is shown in \cref{fig:ltc_signal}.
\begin{table}
    \centering
    \renewcommand{\arraystretch}{1.4}
    \caption{Mismatch, mean absolute error (MAE), and frame shift margin (fsm) for different frame rates over $N$ frames}
    \label{tab:fps_test}
    \footnotesize
        \begin{tabularx}{\columnwidth}{@{}
        >{\centering\arraybackslash}p{0.1\columnwidth}
        >{\centering\arraybackslash}p{0.1\columnwidth}
        >{\centering\arraybackslash}p{0.1\columnwidth}
        >{\centering\arraybackslash}p{0.15\columnwidth}
        >{\centering\arraybackslash}p{0.1\columnwidth}
        >{\centering\arraybackslash}X
        @{}}
        \toprule
        \textbf{FPS} & \(\mathbf{t_{max\,shift}}\) & \(\mathbf{N}\) & \textbf{Mean} & \textbf{MAE} & \textbf{FSM} \\
        \midrule
        24 & $28.8\overline{3}$\,ms & 24    & -1.88\,ms   & 4.88\,ms        & 19.83\,ms  \\          
        25 & 20.00\,ms              & 25    & -5.92 ms    & 7.2\,ms       & 1.24\,ms \\
        30 & $16.\overline{6}$\,ms  & 30    & 6.43 ms    & 7.03\,ms       & 1.18\,ms \\
        \bottomrule
        \end{tabularx}
        \vspace{-4mm}
\end{table}

The comparison between \ac{ltc} signal and the \ac{utc} time reference exhibits a maximum mean offset of up to \qty{6.43}{\milli\second} with a standard deviation of \qty{7.03}{\milli\second} at 30 \ac{fps}.
Relating the results in \cref{tab:fps_test} to the expected RMS timing uncertainty of \qty{6.8}{\milli\second} derived in \cref{eq:theoretical_deviation} shows that the observed results lie within the expected uncertainty of the evaluation \change{setup. Across} all evaluated frame rates, the \ac{fsm} remains greater than \qty{1.18}{\milli\second}, indicating sufficient margin to avoid frame-index ambiguity. As specified in the \ac{smpte} standard~\cite{SMPTEST12-1}, \ac{ltc} encodes time on a per-frame basis and cannot guarantee temporal alignment below half the duration of one frame, as no direct synchronization exists between the camera and the \ac{tcg}.
In such cases, the decoding software assigns the nearest \ac{ltc} timestamp to each frame, which induces bounded timing deviations.
Overall, all tested frame rates satisfy the temporal accuracy requirements typically expected in professional video applications~\cite{SMPTEST12-1}.
The results indicate consistent and stable synchronization of the generated \ac{ltc} signal to the \ac{utc} timebase under the evaluated conditions.
\subsection{Timing Stability}
Applying the timing model described in Section \ref{sec:timing_analysis}, the achievable synchronization duration without active \ac{gnss} correction was evaluated for common video frame rates. For a frame rate of 24\,\ac{fps}, the system remains time-aligned for up to \qty{682}{\second}. At 25\,\ac{fps}, the corresponding upper bound is \qty{654}{\second}. The shortest synchronization interval occurs at 30\,\ac{fps}, where time alignment is maintained for up to \qty{543}{\second}, equivalent to approximately \qty{9}{\minute}, before a potential frame-level timestamp shift may occur. These limits are dominated by the relative frequency error of the internal crystal oscillator rather than by timer discretization effects or computation latency. The deterministic timing deviation introduced by frame discretization remains negligible in comparison. The achievable synchronization duration can be substantially extended by replacing the internal crystal oscillator with a temperature-compensated crystal oscillator. \change{This improvement would come at the cost of increased power consumption and system complexity.}
\subsection{Power Consumption}
The power analysis presented in Section~\ref{sec:timing_analysis} yields the current consumption values summarized in \cref{tab:power_consumption}.

\begin{table}[h]
    \centering
    \renewcommand{\arraystretch}{1.2}
    \caption{Power consumption of active components at \qty{1.8}{\volt}}
    \label{tab:power_consumption}
    \footnotesize
    \begin{tabularx}{\columnwidth}{@{}
        >{\centering\arraybackslash}X
        >{\centering\arraybackslash}X
        >{\centering\arraybackslash}X
        @{}}
        \toprule
        \textbf{Component} & \textbf{Part Name} & \textbf{Power Consumption} \\
        \midrule
        GNSS Module             & MAX-M10S  & \qty{30.92}{\milli\watt} \\
        Microcontroller         & NORA-B101 & \qty{3.74}{\milli\watt}  \\
        Output Buffer           & OPA387    & \qty{730.8}{\micro\watt} \\
        Digital Potentiometer   & MAX5395  & \qty{26.52}{\micro\watt} \\
        \midrule
        \textbf{Full system} & & \textbf{\qty{35.37}{\milli\watt}} \\
        \bottomrule
    \end{tabularx}
\end{table}

The total system power consumption at \qty{1.8}{\volt} amounts to \qty{35.37}{\milli\watt}. A current draw of \qty{12.88}{\milli\ampere} at \qty{3}{\volt} corresponds to an average buck-converter efficiency of \qty{91.5}{\percent}, which is consistent with the datasheet specifications. Assuming a nominal capacity of \qty{1000}{\mAh} for standard AAA alkaline batteries~\cite{aaa_alkaline_mikhayov_2012}, the system achieves an operating time of up to \qty{75}{\hour}.

\section{Conclusion}\label{sec:conclusion}

\change{This work presents a compact and lightweight \acl{tcg} that generates \ac{ltc} signals for video timestamping directly from \ac{gnss}-derived time.
In combination with \ac{gnss}-enabled body-attached sensor nodes, this approach enables precise temporal alignment of video recordings and sensor data without requiring direct coordination between devices. Cameras and sensor nodes operate independently and align their data streams using a common, satellite-provided time reference.
Therefore, the camera and sensor nodes can be powered on immediately before the activity of interest and synchronized directly to the \ac{gnss} time base. This reduces active time and overall power consumption across all system components, enabling smaller batteries and lighter hardware while improving usability in field deployments.}
A detailed timing analysis quantified the individual contributions of timer discretization, computation latency, and oscillator frequency drift.
\change{The results indicate that discretization-induced deviations on the \ac{tcg} side are negligible compared to the crystal oscillator's inherent frequency error.
Furthermore, with a worst-case frame shift margin of \qty{1.18}{\milli\second} at 30\,fps, the system is well within the permissible half-frame tolerance defined by the \ac{ltc} specification.}

\change{When loosing \ac{gnss} synchronization the \ac{tcg} relies on the internal clock for \ac{ltc} timecode generation.
In this case, temporal alignment is maintained for a minimum of \qty{543}{\second} (\qty{9}{\minute}) at \qty{30}{\fps}. After this period, a frame-level timestamp shift may occur, introducing a one-frame temporal offset between the sensor data and the video stream.
With an average power consumption of \qty{35.37}{\milli\watt}, the system achieves a runtime of up to \qty{75}{\hour} when powered from two standard AAA alkaline batteries.}

Future work will integrate the \ac{tcg} into a sports data collection scenario to evaluate its performance and usability as a visual reference for data analysis. Additionally, a more precise reference clock will be explored to extend offline synchronization time.



\begin{acronym}

\acro{soa}[SoA]{State-of-the-Art}
\acro{hmi}[HMI]{Human-Machine Interface}
\acro{2d}[2D]{two-dimensional}
\acro{pc}[PC]{personal computer}
\acro{ee}[EE]{electrical engineering}

\acro{iot}[IoT]{Internet-of-Things}
\acro{fpga}[FPGA]{Field Programmable Gate Arrays}
\acro{cots}[COTS]{common off-the-shelf}
\acro{mcu}[MCU]{microcontrollers}
\acro{pll}[PLL]{phase-locked loop}
\acro{dma}[DMA]{Direct memory access}
\acro{opamp}[op-amp]{operational amplifiers}
\acro{fifo}[FIFO]{First In, First Out}
\acro{led}[LED]{light-emitting diode}
\acro{adc}[ADC]{analog-digital converter}
\acro{soc}[SoC]{system-on-chip}
\acro{ic}[IC]{integrated circuit}
\acro{rf}[RF]{radio frequency}
\acro{ble}[BLE]{Bluetooth low energy}
\acro{lut}[LUT]{Lookup Table}
\acro{tia}[TIA]{transimpedance amplifier}
\acro{wcet}[WCET]{Worst Case Execution Time}
\acro{uart}[UART]{Universal Asynchronous Receiver/Transmitter}
\acro{pwm}[PWM]{pulse-width modulation}
\acro{lptim}[LPTIM]{Low Power Timer}
\acro{fpu}[FPU]{Floating Point Unit}
\acro{gpio}[GPIO]{General-Purpose Input/Output}
\acro{simd}[SIMD]{Single Instruction Multiple Data}
\acro{i2c}[I2C]{InnInter-Integrated Circuit}
\acro{nvic}[NVIC]{Nested Vector Interrupt Controller}
\acro{spi}[SPI]{Serial Peripheral Interface}
\acro{sdo}[SDO]{Serial Data Out}
\acro{sdi}[SDI]{Serial Data IN}
\acro{sck}[SCK]{Serial Clock}
\acro{spst}[SPST]{Single-Pole Single-Throw}
\acro{pdm}[PDM]{Pulse Density Modulation}
\acro{dsp}[DSP]{Digital Signal Processing}
\acro{rtos}[RTOS]{Real-Time Operating Systems}
\acro{gpio}[GPIO]{General-Purpose Input-Output}
\acro{imu}[IMU]{Inertial Measurement Unit}
\acro{dma}[DMA]{Direct Memory Access}
\acro{cmsis}[CMSIS]{Common Microcontroller Software Interface Standard}
\acro{os}[OS]{operating system}
\acro{isa}[ISA]{Instruction Set Architecture}
\acro{cpu}[CPU]{Central Processing Unit}

\acro{iis}[IIS]{Integrated Systems Laboratory}
\acro{led}[LED]{Light-Emitting Diode}
\acro{ic}[IC]{Integrated Circuit}
\acro{ltc}[LTC]{Linear Timecode}
\acro{gps}[GPS]{Global Positioning System}
\acro{gnss}[GNSS]{Global Navigation Satellite System}
\acro{smpte}[SMPTE]{Society of Motion Picture and Television Engineers}
\acro{fps}[fps]{frames per second}
\acro{bgfb}[bGFb]{Binary Group Flag Bits}
\acro{bcd}[BCD]{Binary Coded Decimal}
\acro{eth}[ETH]{Eidgenössische Technische Hochschule}
\acro{utc}[UTC]{Coordinated Universal Time}
\acro{usb}[USB]{Universal Serial Bus}
\acro{mcu}[MCU]{Microcontroller Unit}
\acro{uart}[UART]{Universal Asynchronous Receiver/Transmitter}
\acro{i2c}[I2C]{Inter-Integrated Circuit}
\acro{opamp}[Op-Amp]{Operational Amplifier}
\acro{trs}[TRS]{Tip-Ring-Sleeve}
\acro{mmcx}[MMCX]{Micro-Miniature-Coaxial}
\acro{tcg}[TCG]{Timecode Generator}
\acro{gpio}[GPIO]{General-Purpose Input/Output}
\acro{cpu}[CPU]{Central Processing Unit}
\acro{nmea}[NMEA]{National Marine Electronics Association}
\acro{dc}[DC]{Direct Current}
\acro{pcb}[PCB]{Printed Circuit Board}
\acro{rc}[RC]{Resistor-Capacitor}
\acro{ppm}[ppm]{parts per million}
\acro{pbl}[PBL]{Center for Project-Based Learning}
\acro{bmc}[BMC]{Biphase Mark Code}
\acro{smpte}[SMPTE]{Society of Motion Picture and Television Engineers}
\acro{ntp}[NTP]{Network Time Protocol}

\acro{ppg}[PPG]{photoplethysmogram}

\acro{fft}[FFT]{Fast Fourier Transform}
\acro{snr}[SNR]{signal-to-noise ratio}
\acro{dsp}[DSP]{digital signal processing}
\acro{iir}[IIR]{infinite impulse response}

\acro{tcxo}[TCXO]{Temperature-Compensated Crystal Oscillator}

\acro{sd}[SD]{Standard Deviation}
\acro{rmse}[RMSE]{Root Mean Square Error}
\acro{mae}[MAE]{Mean Absolute Error}
\acro{maxae}[MaxAE]{Maximum Absolute Error}
\acro{95ae}[95AE]{95th-percentile Absolute Error}
\acro{99ae}[99AE]{99th-percentile Absolute Error}
\acro{fsm}[FSM]{Frame Shift Margin}

\acro{ae}[AE]{absolute error}
\acro{me}[ME]{mean error}

\acro{pbl}[PBL]{Project-Based Learning}
    
\end{acronym}

\section*{ACKNOWLEDGMENT}
This work was supported in part by the CHIST-ERA project ”SNOW” (Grant 209675) and the Ambizione Grant 233457.

\bibliographystyle{IEEEtranDOI} 

\bibliography{
    bib/sas,
    bib/pleisch
}

\end{document}